\documentclass[11pt]{article}
\usepackage{amsmath}
\usepackage{amsfonts}
\usepackage{amssymb}
\def\bbox#1{{\mathbf{#1}}}
\newcommand{\dbar}{\bar{\partial}}
\newcommand{\D}{\mathcal{D}}

\newcommand{\be}{\begin{equation}}
\newcommand{\ee}{\end{equation}}
\newcommand{\bea}{\begin{eqnarray}}
\newcommand{\eea}{\end{eqnarray}}
\newcommand{\beaa}{\begin{eqnarray*}}
\newcommand{\eeaa}{\end{eqnarray*}}

\newcommand{\nn}{\nonumber}
\begin{document}
%%%%%%%%%%%%%%%%%%%%%%%%%%%%%%%%%%%%%%%%%%%%%%%%%%%%%%%%%%%%
\title{Projective differential geometry of higher 
reductions of the two-dimensional Dirac equation}
\author{{\Large L.V. Bogdanov}  \\
     Landau Institute for Theoretical Physics RAS\\
     Kosygin str. 2,
    119334 Moscow, Russia\\
    e-mail: {\tt leonid@landau.ac.ru}\\
     and \\
     {\Large E.V. Ferapontov} \\
     Department of Mathematical Sciences \\
    Loughborough University \\
    Loughborough, Leicestershire LE11 3TU \\
    United Kingdom \\
    e-mail: {\tt E.V.Ferapontov@lboro.ac.uk}}
\date{}
\maketitle

\newtheorem{theorem}{Theorem}

\pagestyle{plain}

\maketitle

\begin{abstract}

We investigate reductions of the two-dimensional Dirac equation 
imposed by the requirement of the existence of a differential operator $D_n$
of order $n$ mapping its 
eigenfunctions to adjoint eigenfunctions. For first order operators these
reductions (and multi-component analogs thereof) 
lead to the Lame equations descriptive of orthogonal coordinate systems. 
Our main observation is that
$n$-th order reductions coincide with the projective-geometric `Gauss-Codazzi' 
equations governing special classes of line congruences in the projective space 
$P^{2n-1}$, which is the projectivised kernel of $D_n$.
In the second order case  this leads to the theory of $W$-congruences  in $P^3$ 
which belong to a linear complex, while
the third order case  corresponds to isotropic congruences in $P^5$.
Higher  reductions are compatible with  odd-order flows of the 
Davey-Stewartson hierarchy. All these flows preserve the kernel
$D_n$, thus defining nontrivial geometric evolutions of line congruences.

Multi-component generalizations are also discussed. The correspondence
between geometric picture and the theory of integrable systems is established; the
definition of the class of reductions and all geometric objects
in terms of the multicomponent KP hierarchy is presented. Generating forms
for reductions of arbitrary order are constructed.

\bigskip


\bigskip

\end{abstract}

\newpage
\section{Introduction}
The two-dimensional Dirac equation,
\begin{equation}
\partial_x\Psi_2=\beta \Psi_1, ~~~ \partial_y\Psi_1=\gamma \Psi_2,
\label{Dirac}
\end{equation}
plays an important role in differential geometry and
mathematical physics,  
arising, in particular,  as the Lax operator of the Davey-Stewartson (DS)
hierarchy. Our main goal is to investigate  the class of its reductions
(that is,  differential constraints on the potentials $\beta(x, y)$ and 
$\gamma(x, y)$, possibly nonlocal), which are compatible with odd order flows 
of the DS hierarchy
(the modified Veselov-Novikov system being the first nontrivial flow among them
\cite{Bogdanov}). The simplest representative of this class of reductions
(in one-component case) is probably the BKP hierarchy \cite{BKP}.
Reductions describing Veselov-Novikov equation and modified Veselov-Novikov
equation \cite{VesNov,GM,Bogdanov}
also belong to the class  we study.
In terms of the $\dbar$-dressing method these reductions manifest themselves
as special linear conditions on the kernel of the nonlocal $\dbar$-problem,
and it was shown  in \cite{Zakharov1} that Lame equations
describing N-orthogonal coordinate systems can be obtained in this way. 
In the work \cite{ZM} a general class of conditions
on the kernel of the nonlocal $\dbar$-problem was described.

The general approach we adopt here is to require the existence of a linear
differential operator $D_n$ of order $n$ which
maps solutions of (\ref{Dirac}) to solutions of the adjoint problem,
\begin{equation}
\partial_x\Psi^*_2=\gamma \Psi^*_1, ~~~ \partial_y\Psi^*_1=\beta \Psi^*_2.
\label{Diracadj}
\end{equation}
In this direct form the method was probably first proposed in   
\cite{Schief}.  Explicitly, we require 
\begin{equation}
\Psi^*_1={c}_1\partial_x^n\Psi_1+ ..., ~~~~
\Psi^*_2={c}_2\partial_y^n\Psi_2+ ...,
\label{D}
\end{equation}
where ${c}_i=\text{const}$ and dots denote an arbitrary linear combination
of the terms $\partial_x^k\Psi_1$ ans $\partial_y^k\Psi_2$, $\ k=0, 1, ..., n-1$.
Notice that the requirement of $\Psi^*_1, \ \Psi^*_2$ 
being adjoint eigenfunctions imposes strong constraints on the coefficients
in (\ref{D}), specifying them almost uniquely up to a certain natural equivalence. 
The zero order case, 
$$
\Psi^*_1={c}_1\Psi_1, ~~~~
\Psi^*_2={c}_2\Psi_2,
$$
implies the well-known reduction
$$
{c}_1\gamma={c}_2\beta.
$$
In the case of first order operators,
$$
\Psi^*_1={c}_1\partial_x\Psi_1+{c}_2\beta \Psi_2, ~~~~
\Psi^*_2={c}_2\partial_y\Psi_2+{c}_1\gamma \Psi_1,
$$
one obtains the reduction
$$
{c}_1\gamma_x+{c}_2\beta_y=0.
$$

Our main observation is that reductions of higher order are intimately 
connected with projective differential geometry. Moreover, the venue of this geometry is
the projectivised kernel $P^{2n-1}$ of the operator $D_n$. 
For instance, the case of $D_2$ is characterised by
\begin{equation}
\begin{array}{c}
\Psi_1^*=c_1(\partial^2_x\Psi_1-U\Psi_1)+c_2(\beta\partial_y\Psi_2-
\beta_y\Psi_2), \vspace{1\jot}
\\
\Psi_2^*=c_2(\partial^2_y\Psi_2-V\Psi_2)+c_1(\gamma\partial_x\Psi_1-
\gamma_x\Psi_1),
\end{array}
\label{D2}
\end{equation}
where the potentials $\beta, \ \gamma$ and the `nonlocalities' $U, \ V$
satisfy the reduction equations
\begin{equation}
\begin{array}{c}
c_1(\gamma_{xx}-U\gamma)=c_2(\beta_{yy}-V\beta),
\vspace{2\jot}\\
U_y=3\gamma_x\beta+\gamma\beta_x, 
\vspace{1\jot}\\
V_x=3\beta_y\gamma+\beta\gamma_y.
\end{array}
\label{red2}
\end{equation}
The 4-dimensional kernel of $D_2$ is defined by the equations
\begin{equation}
\begin{array}{c}
\partial_x\Psi_2=\beta \Psi_1, ~~~ \partial_y\Psi_1=\gamma \Psi_2, 
\vspace{2\jot}\\
\partial^2_x\Psi_1=U\Psi_1-\frac{c_2}{c_1}(\beta\partial_y\Psi_2-
\beta_y\Psi_2), 
\vspace{1\jot}\\
\partial^2_y\Psi_2=V\Psi_2-\frac{c_1}{c_2}(\gamma\partial_x\Psi_1-
\gamma_x\Psi_1), 
\end{array}
\label{ker2}
\end{equation}
which, as we point out in section 2,  coincide with the standard Wilczynski-type
 equations of a 
W-congruence in $P^3$ with two focal surfaces $\Psi_1$ and $\Psi_2$,
referred to conjugate coordinates $x$ and $y$. Moreover, 
this congruence lies in a linear complex which can be constructed as follows. 
Let us introduce a potential $S$ by the formulae
$$
S_x=\Psi_1 \Phi_1^*-\Psi_1^*\Phi_1, ~~~ S_y=\Psi_2 \Phi_2^*-\Psi_2^* \Phi_2,
$$ 
which  explicitly integrate to a skew-symmetric bilinear form S on the space of solutions
of (\ref{Dirac}),
\begin{equation}
S(\Psi, \  \Phi)= c_1(\Psi_1\partial_x \Phi_1-\Phi_1\partial_x \Psi_1)+
 c_2(\Psi_2\partial_y \Phi_2-\Phi_2\partial_y \Psi_2).
\label{complex}
\end{equation}
The restriction of $S$ to the kernel of $D_2$ 
is the  invariant skew-symmetric bilinear form  
defining  a linear complex of lines in $P^3$ (recall that a linear 
complex  in $P^3$ is given by one  linear equation in Pl\"ucker coordinates, that is, 
by a $4\times 4$ skew-symmetric matrix). As we demonstrate in section 2, 
the congruence (\ref{ker2}) lies in the linear complex defined by $S$. 
The reduction equations
(\ref{red2}) are nothing but the corresponding projective `Gauss-Codazzi' 
equations.

Similarly,  the general form of $D_3$ is
\begin{equation}
\begin{array}{c}
\Psi_1^*=c_1(\partial_x^3\Psi_1-2V\partial_x\Psi_1-V_x\Psi_1)
+c_2(\beta\partial_y^2\Psi_2-\beta_y\partial_y\Psi_2
+(\beta_{yy}-2V\beta)\Psi_2),
\vspace{1\jot}\\
\Psi_2^*=c_2(\partial_y^3\Psi_2-2W\partial_y\Psi_2-W_y\Psi_2)
+c_1(\gamma\partial_x^2\Psi_1-\gamma_x\partial_x\Psi_1
+(\gamma_{xx}-2W\gamma)\Psi_1),
\end{array}
\label{D3}
\end{equation}
where the potentials $\beta, \gamma$ and the nonlocalities $V$, $W$ satisfy
the third order reduction equations
\begin{equation}
\begin{array}{c}
c_1(\beta_{yyy}-2\beta_yW-\beta W_y)+
c_2(\gamma_{xxx}-2\gamma_x V-\gamma V_x)=0,
\vspace{2\jot}\\
W_x=2\gamma\beta_y+\beta\gamma_y,
\vspace{1\jot}\\
V_y=2\beta\gamma_x+\gamma\beta_x.
\end{array}
\label{red3}
\end{equation}
The kernel of $D_3$ is six-dimensional, defined by the 
equations
\begin{equation}
\begin{array}{c}
\partial_x\Psi_2=\beta \Psi_1, ~~~ \partial_y\Psi_1=\gamma \Psi_2, 
\vspace{2\jot}\\
\partial_x^3\Psi_1=2V\partial_x\Psi_1+V_x\Psi_1
-\frac{c_2}{c_1}(\beta\partial_y^2\Psi_2-\beta_y\partial_y\Psi_2
+(\beta_{yy}-2V\beta)\Psi_2),
\vspace{1\jot}\\
\partial_y^3\Psi_2=2W\partial_y\Psi_2+W_y\Psi_2
-\frac{c_1}{c_2}(\gamma\partial_x^2\Psi_1-\gamma_x\partial_x\Psi_1
+(\gamma_{xx}-2W\gamma)\Psi_1),
\end{array}
\label{ker3}
\end{equation}
which, as we demonstrate in section 3, also give rise to a line congruence with the 
two focal surfaces 
$\Psi_1$ and $\Psi_2$. Moreover, in this case one can introduce a potential
$S$ by the formulae
$$
S_x=\Psi_1 \Psi_1^*, ~~~ S_y=\Psi_2 \Psi_2^*,
$$ 
which  explicitly integrate to a quadratic form S on 
the space of solutions
of (\ref{Dirac}),
\begin{equation}
S(\Psi, \ \Psi)= c_1(\Psi_1\partial_x^2\Psi_1-\frac{1}{2}(\partial_x\Psi_1)^2-
V\Psi_1^2)+
 c_2(\Psi_2\partial_y^2\Psi_2-\frac{1}{2}(\partial_y\Psi_2)^2-
W\Psi_2^2).
\label{S}
\end{equation}
 In the case $c_1=-c_2=1$, the restriction of $S$ to the kernel of $D_3$ defines the
invariant symmetric scalar product of the signature 
$(3, 3)$. The corresponding
congruence is isotropic with respect to $S$, thus coinciding with the Pl\"ucker image 
of a surface in $P^3$. Reduction equations (\ref{red3}) are nothing but
the projective
`Gauss-Codazzi' equations of surfaces in $P^3$. The necessary 
information on Wilczynski's approach to surfaces in $P^3$ and their Pl\"ucker 
images in $P^5$ is
included in the Appendix A. 

In the case $c_1=c_2=1$, the linear system 
(\ref{ker3}) is descriptive of a surface in Lie sphere geometry (in hexaspherical
representation), equations (\ref{red3}) being the corresponding Lie-geometric
`Gauss-Codazzi' equations \cite{Fer4}. The quadratic form $S$ is in this case
 of the signature $(4, 2)$,
defining the Lie-invariant scalar product on the kernel of $D_3$.

All higher reductions are invariant under B\"acklund transformations of the 
Darboux-Levi type \cite{Matveev}. In the first order case this was pointed out in
\cite{Schief}. In the case of third order reductions 
the corresponding 
Darboux-Levi transformations  coincide with transformations W (Appendix B).

This approach obviously carries over to the multi-component linear problem 
\begin{equation}
\partial_i\Psi_j=\beta_{ij} \Psi_i,
\label{Darboux}
\end{equation}
$i, j=1, ..., N;$ here the potentials $\beta_{ij}$ must satisfy 
the compatibility conditions
$\partial_k\beta_{ij}=\beta_{ik}\beta_{kj}$.
Introducing the adjoint linear problem
\bea
\partial_i\Psi^*_j=\beta_{ji} \Psi^*_i,
\label{*Darboux}
\eea
we define n-th order reductions by  requiring the existence of an operator 
$D_n$ of order $n$, mapping eigenfunctions
$\Psi_i$ to the corresponding adjoint eigenfunctions,
\bea
\Psi^*_i={c}_i\partial_i^n\Psi_i+..., ~~~~ {c}_i=\text{const}.
\label{trans}
\eea
In the zero order case  we have
$$
\Psi^*_i={c}_i \Psi_i, 
$$
implying the familiar `Egorov' reduction
$$
{c}_j \beta_{ij}={c}_i \beta_{ji}.
$$
First order reductions are defined by
$$
\Psi^*_i={c}_i\partial_i\Psi_i+\sum_{k\ne i} {c}_k
\beta_{ik}\Psi_k,
$$
leading to the 'Lame' equations
$$
{c}_i\partial_i\beta_{ji}+{c}_j\partial_j\beta_{ij}+
\sum_{k\ne i, j} {c}_k \beta_{ik}\beta_{jk}=0
$$
which are descriptive of $N$-orthogonal coordinate systems. 
In this form, first order reductions of 
the system (\ref{Darboux}) were discussed by Schief \cite{Schief}, see also
\cite{Zakharov1}, \cite{Zakharov2} for an alternative approach.

In the multi-component situation, all higher  reductions also 
have a clear differential-geometric interpretation, 
governing special Laplace sequences (of congruences lying in linear complexes, isotropic 
congruences, etc; see sections 2 and 3 for 
explicit formulae and geometric discussion).

The class of reductions of the form (\ref{D}) has a very natural interpretation
in terms of the multicomponent KP hierarchy. In fact, 
this work started from the observation
that `Gauss-Codazzi' equations (\ref{red3}) of surfaces in $P^3$ belong to a special
class of  reductions
of Davey-Stewartson hierarchy with odd times. This oservation has given an impetus
to the study of connections of this class of reductions with projective geometry,
the first results of which are presented in this work. 
 
Multicomponent KP hierarchy corresponds to the dynamics of the multicomponent
Grassmannian defined by the loop group \cite{Sato},
\cite{Segal}, \cite{Kac}. Let  the Grassmannian point  $W(\lambda)$
be realized as a space of vector functions (rows of  length $N$)
on the unit circle  $\bbox{S}$ in the complex plane $\mathbb{C}$, 
and the dual point  $W^*(\lambda)$ as a space of 
functions taking their values
in the space of columns of height $N$,
\bea
\oint W(\lambda)W^*(\lambda)d\lambda=0.
\label{dual}
\eea 
The loop group ${\Gamma^+}^N$, represented here by  $N\times N$ diagonal  matrix
functions $g(\lambda)$ on $\bbox{S}$, $g_{ij}(\lambda)=\delta_{ij}g_i(\lambda)$,
$g_i(\lambda)\in \Gamma^{+}$ (i.e., functions $g_i(\lambda)$ are analytic functions
in the unit disc $\bbox{D}$ having no zeroes), defines 
a dynamics (deformations) on the Grassmannian,
\bea
W(g;\lambda)=W_0(\lambda)g^{-1}(\lambda),\quad
W^*(g;\lambda)=g(\lambda)W^*_0(\lambda),
\label{dynamics}
\eea
which evidently preserves the duality (\ref{dual}).
If we introduce the standard parameterization of ${\Gamma^+}^N$ in terms
of an infinite number of variables (`times' of the hierarchy),
\bea
g_i(\lambda)=g_i(\lambda,\bbox{x})=\exp
\left(\sum_{n=1}^{\infty}x_{i(n)}\lambda^n\right),
\label{parameterization}
\eea
the  points of the Grassmannian will depend on infinite number of `times', leading 
to the standard picture of the hierarchy in terms of PDEs. 
It is easy to derive linear equations (\ref{Darboux}) and dual equations
(\ref{*Darboux})
in this context (see, e.g., \cite{LVBbook}).
The variables
$x_i=x_{i(1)}$ correspond to the variables $x_i$  introduced above.

The standard reduction of KP hierarchy describing, in  scalar case, the n-th 
Gelfand-Dickey hierarchy (n=2 corresponds to KdV hierarchy),
is the condition
\bea
\lambda^n W(\lambda)\subset W(\lambda),
\eea
which is preserved by the dynamics and leads to the stationarity
of the point of the Grassmannian with respect to some `time' of the hierarchy. 
In this sense
conditions of this type always lead to dimensional reduction 
(from a hierarchy of (2+1)-dimensional integrable equations to the 
hierarchy of (1+1)-dimensional integrable equations). The important feature
of the class of reductions we study is that these reductions 
{\em do not reduce the dimension}, and the corresponding integrable 
systems are (2+1)-dimensional. These reductions are consistent not
with the full dynamics  given by ${\Gamma^+}$, rather than only
the dynamics connected with its subgroup
defined by the condition 
\bea
g(\lambda)g(-\lambda)=I.
\label{subgroup}
\eea
In terms of  the times  (\ref{parameterization}),
this condition just means that even times are equal to zero,
and only evolution with respect to odd times is considered.

In this case, in addition to a pair of dual Grassmannian points,
we also consider a dual pair 
${W^*}^{\text{t}}(g,-\lambda)$ and $W^{\text{t}}(g;-\lambda)$
(`t' is for transposed).
It is easy to see that the dynamics 
(\ref{dynamics})
defined by the subgroup
(\ref{subgroup}) is identical on the spaces 
$W(g;\lambda)$ and ${W^*}^{\text{t}}(g,-\lambda)$
($W^*(g;\lambda)$ and ${W}^{\text{t}}(g,-\lambda)$, respectively).
 
Thus, conditions of the type
\bea
{W^*}^{\text{t}}(g;-\lambda)F(\lambda) \subset W(g;\lambda),
\label{reduction0}
\eea
where $F(\lambda)$ is a diagonal matrix, $F_{ij}=\delta_{ij}F_i(\lambda)=
\delta_{ij}c_i\lambda^n$, are invariant in the dynamics, and define
reductions of the initial system. This is exactly the class of reductions (\ref{D})
we have introduced above , and it is a rather straightforward technical
step to derive transformation (\ref{D}) from the condition (\ref{reduction0}).

Solutions to multicomponent KP hierarchy can be constructed 
using the $\dbar$-dressing method \cite{ZM85}. The spaces $W(\bbox{x},\lambda)$
and $W^*(\bbox{x},\lambda)$ correspond in this case to  spaces of solutions
of the nonlocal $\dbar$-problem and the dual problem 
(see the explicit construction
in \cite{LVBbook}). The class of reductions (\ref{reduction0}) corresponds in this case
to very simple conditions on the kernel of the nonlocal matrix
$\dbar$-problem,
$$
R^{\text{t}}(-\mu,-\lambda)=F(\mu)R(\lambda,\mu)F^{-1}(\lambda).
$$
Conditions of this type were considered in \cite{GM},\cite{Bogdanov},
\cite{Zakharov1,ZM,Zakharov2}.

%%%%%%%%%%%%%%%%%%%%%%%%%%%%%%%%%%%%%%%%%%%%%%%%%%%%%%%%%%%%%%%%%%%%%%
\section{Second order reductions and congruences  in $P^3$ lying in  linear complexes}
%%%%%%%%%%%%%%%%%%%%%%%%%%%%%%%%%%%%%%%%%%%%%%%%%%%%%%%%%%%%%%%%%%%%%%%
\subsection{Two-component case}
%%%%%%%%%%%%%%%%%%%%%%%%%%%%%%%%%%%%%%%%%%%%%%%%%%%%%%%%%%%%%%%%%%%%%%%%
Under the reduction conditions (\ref{red2}), the linear system
(\ref{ker2}),
$$
\begin{array}{c}
\partial_x\Psi_2=\beta \Psi_1, ~~~ \partial_y\Psi_1=\gamma \Psi_2, 
\vspace{2\jot}\\
\partial^2_x\Psi_1=U\Psi_1-\frac{c_2}{c_1}(\beta\partial_y\Psi_2-
\beta_y\Psi_2), 
\vspace{1\jot}\\
\partial^2_y\Psi_2=V\Psi_2-\frac{c_1}{c_2}(\gamma\partial_x\Psi_1-
\gamma_x\Psi_1), 
\end{array}
$$
which describes the kernel of $D_2$, is compatible of rank 4
(indeed, the solution is completely determined by the values of
$\Psi_1, \ \Psi_2, \ \partial_x\Psi_1, \ \partial_y\Psi_2$ at a fixed point
$x, y$). Therefore, both $\Psi_1(x, y)$ and $\Psi_2(x, y)$ can
be interpreted as 4-component position vectors  of 
two surfaces in $P^3$,  $M_1$ and $M_2$ (written in homogeneous coordinates).
 The first two equations
in (\ref{ker2}) clearly imply that the parametrization $x, y$ is conjugate, 
moreover,  the line $(\Psi_1, \ \Psi_2)$ is
tangential to both $M_1$ and $M_2$ along the $y-$ and $x-$directions, 
respectively. In other words, $M_1$ and $M_2$ are  focal surfaces of the 
line congruence $(\Psi_1, \ \Psi_2)$. Historically, the theory of 
line congruences (that is, 
2-parameter families of lines in $P^3$) has been one of the most popular 
chapters of projective differential geometry, dating back to Monge, Pl\"ucker and 
Kummer. In the majority of geometrical applications the main role is played by 
special congruences known as congruences W (named after Weingarten), which
are characterised by the property that the (projective) second fundamental
forms of the focal surfaces coincide. A direct computation shows that 
second fundamental forms of both focal surfaces $M_1$ and $M_2$ indeed coincide, 
both being conformal to $c_2\beta dx^2-c_1\gamma dy^2$ 
(recall that only the conformal class 
of the second fundamental form is a projective invariant). 

Let us introduce a skew-symmetric scalar product $\{~, ~\}$ on the kernel of $D_2$ 
defined by the bilinear form (\ref{complex}),
\begin{equation}
\{\Psi_1, \ \partial_x\Psi_1\}=1/c_1, ~~~ \{\Psi_2, \ \partial_y\Psi_2\}=1/c_2,
\label{complex1}
\end{equation}
(all other scalar products being zero). Since the two-dimensional plane spanned by
$\Psi_1$ and $\Psi_2$ in the four-dimensional kernel 
of $D_2$ is clearly Lagrangian (indeed, $\{\Psi_1, \ \Psi_2\}=0$),  
the projective line $(\Psi_1, \ \Psi_2)$ belongs to the corresponding linear complex.
The representation of congruences of linear complexes in the form
(\ref{ker2}) can be traced back to the works \cite{Wilczynski1, Wilczynski2}. 
The compatibility conditions (\ref{red2})
are nothing but the corresponding projective `Gauss-Codazzi' equations. 
It should be emphasized that congruences of linear complexes 
constitute a proper subclass of congruences W.

All odd order flows of the DS hierarchy preserve the kernel of $D_2$ and 
the skew-symmetric form $S$.  Thus, they  induce geometric evolutions of congruences of 
linear complexes (so that the  equation of the  complex stays the same in the 
process of evolution). Moreover, these evolutions 
preserve the conjugate parametrization $x, y$ and the projectively invariant functional
$$
\int \beta \gamma \ dxdy,
$$
which is the first nontrivial conservation law of the DS hierarchy.

Similar integrable evolutions  in conformal, affine and Lie 
sphere geometries were a subject of recent publications 
\cite{Kon1, Taimanov, Burstall, Kon2, KonPin, Fer5}.

%%%%%%%%%%%%%%%%%%%%%%%%%%%%%%%%%%%%%%%%%%%%%%%%%%%%%%%%%%%%%%%%%%%%%%
\subsection{N-component case}
In the N-component case (\ref{Darboux}), equations (\ref{D2}) take the form
\begin{eqnarray}
\Psi_i^*=
c_i(\partial^2_i\Psi_i-U_i\Psi_i)
+\sum\limits_{\substack{1\leqslant p \leqslant N,\\ p\neq i}}
c_p(\beta_{ip}\partial_p\Psi_p-(\partial_p\beta_{ip})\Psi_p),
\end{eqnarray}
leading to the reduction  
\begin{equation}
\begin{array}{c}
c_i(\partial^2_i\beta_{ji}-U_i\beta_{ji})-
c_j(\partial^2_j\beta_{ij}-U_j\beta_{ij})
%\nn\\
+\displaystyle\sum\limits_{\substack{1\leqslant p \leqslant N,\\ p\neq i,j}}
c_p(\beta_{ip}\partial_p\beta_{jp}-
\beta_{jp}\partial_p\beta_{ip})=0, 
\vspace{1\jot}\\
\partial_j U_i=3\beta_{ij}\partial_i\beta_{ji}+\beta_{ji}\partial_i\beta_{ij},
\quad i\neq j,
\end{array}
\label{red2N}
\end{equation}
which, for $N=2$, coincides with (\ref{red2}) after the identification
$\beta_{12}=\beta, \ \beta_{21}=\gamma$. The kernel of $D_2$ is defined by the 
linear system
\begin{gather*}
\partial_i\Psi_j=\beta_{ij} \Psi_i,\\
\partial^2_i\Psi_i=U_i\Psi_i
-\sum\limits_{\substack{1\leqslant p \leqslant N,\\ p\neq i}}
\frac{c_p}{c_i}(\beta_{ip}\partial_p\Psi_p-(\partial_p\beta_{ip})\Psi_p),
\end{gather*}
which is compatible of rank $2N$ by virtue of the reduction equations
(\ref{red2N}). Each of the $\Psi_i$'s can thus be regarded as a position vector of an
$N$-dimensional submanifold $M_i$ in $P^{2N-1}$, parametrized by conjugate coordinates
$x_1, ..., x_n, \ \partial_i=\partial/\partial_{x_i}$. Moreover, the line 
 $(\Psi_i, \ \Psi_j)$ is tangential to both $M_i$ and $M_j$.
A simple calculation shows that the second fundamental forms of $M_i$ 
(notice that there should be $N-1$ thereof, where $N-1$ is the codimension 
of $M_i$ in $P^{2N-1}$), are given by 
$$
c_i\beta_{pi}dx_p^2-c_p\beta_{ip}dx_i^2, ~~~ p\ne i.
$$
Each pair of  focal submanifolds (say, $M_i$ and  $M_j$), has one second 
fundamental form `in common', namely, the form 
$c_i\beta_{ji}dx_j^2-c_j\beta_{ij}dx_i^2$. This can be interpreted as a 
multicodimensional analog of  the W-property.

Let us introduce a potential $S$ by the formulae
\begin{equation}
\partial_iS=\Psi_i \Phi_i^*-\Psi_i^*\Phi_i,
\label{Sdiffeven}
\end{equation} 
which  explicitly integrate to a skew-symmetric bilinear form S on the space of solutions
of (\ref{Darboux}),
\begin{equation}
S(\Psi, \  \Phi)=\sum_{i=1}^{N} c_i(\Psi_i\partial_i \Phi_i-\Phi_i\partial_i\Psi_i).
\label{complexN}
\end{equation}
The restriction of $S$ to the kernel of $D_2$ is the invariant 
skew-symmetric bilinear
form  corresponding to the skew-symmetric scalar product
\begin{equation}
\{\Psi_i, \ \partial_i\Psi_i\}=1/c_i,
\label{complex1N}
\end{equation}
(all other scalar products being zero). 
Since the $N$-dimensional subspace spanned by $\Psi_i$
in the $2N$-dimensional kernel of $D_2$ is clearly Lagrangian 
(indeed, $\{\Psi_i, \ \Psi_j\}=0$),
each of the congruences $(\Psi_i, \ \Psi_j)$ 
belongs to one and the same linear complex defined
by $S$.

To the best of our knowledge, the geometric object consisting of
N submanifolds of codimension $N-1$ in $P^{2N-1}$, connected  
by congruences which belong to a linear complex, has not been discussed before.

%%%%%%%%%%%%%%%%%%%%%%%%%%%%%%%%%%%%%%%%%%%%%%%%%%%%%%%%%%%%%%%%%%%%%%%%%%%%%%%%%%%%%%
\section{Third order reductions and isotropic congruences in $P^5$}
%%%%%%%%%%%%%%%%%%%%%%%%%%%%%%%%%%%%%%%%%%%%%%%%%%%%%%%%%%%%%%%%%%%%%%%%%%%%%%%%%%%%%
\subsection{Two-component case}
%%%%%%%%%%%%%%%%%%%%%%%%%%%%%%%%%%%%%%%%%%%%%%%%%%%%%%%%%%%%%%%%%%%%%%%%%%%%%%%%%%%%%
Under the reduction equations (\ref{red3}), the linear system (\ref{ker3}),
\begin{equation*}
\begin{array}{c}
\partial_x\Psi_2=\beta \Psi_1, ~~~ \partial_y\Psi_1=\gamma \Psi_2, 
\vspace{2\jot}\\
\partial_x^3\Psi_1=2V\partial_x\Psi_1+V_x\Psi_1
-\frac{c_2}{c_1}(\beta\partial_y^2\Psi_2-\beta_y\partial_y\Psi_2
+(\beta_{yy}-2V\beta)\Psi_2),
\vspace{1\jot}\\
\partial_y^3\Psi_2=2W\partial_y\Psi_2+W_y\Psi_2
-\frac{c_1}{c_2}(\gamma\partial_x^2\Psi_1-\gamma_x\partial_x\Psi_1
+(\gamma_{xx}-2W\gamma)\Psi_1),
\end{array}
%\label{ker3}
\end{equation*}
describing the kernel of $D_3$, is compatible of rank 6
(indeed, the solution is completely determined by the values of
$\Psi_1, \ \Psi_2, \ \partial_x\Psi_1, \ \partial_y\Psi_2,
\ \partial_x^2\Psi_1, \ \partial_y^2\Psi_2$ at a fixed point
$x, y$). Therefore, both $\Psi_1(x, y)$ and $\Psi_2(x, y)$ can
be interpreted as  position vectors  of 
two surfaces in $P^5$,  $M_1$ and $M_2$.
Like in  the second order case,  the first two equations
in (\ref{ker3})  imply that the parametrization $x, y$ is conjugate, 
moreover,  the line $(\Psi_1, \ \Psi_2)$ is
tangential to both $M_1$ and $M_2$ along the $y-$ and $x-$directions, 
respectively. In other words, $M_1$ and $M_2$ are   focal surfaces of the 
line congruence $(\Psi_1, \ \Psi_2)$. In the 
case $c_1=-c_2=1$, the quadratic form
$S$ given by (\ref{S}) defines an invariant symmetric scalar product of 
the signature (3, 3) on the kernel of $D_3$ (see Appendix A). Moreover, 
 the congruence $(\Psi_1, \Psi_2)$ is isotropic with respect to $S$.
This implies that  $M_1$ and $M_2$ are  Pl\"ucker images of asymptotic tangents
 to a surface in $P^3$. The passage from a surface in $P^3$ to its Pl\"ucker 
image in $P^5$ is a classical projective-geometric construction discussed in detail in 
\cite{Bol, Blaschke, Finikov37, Finikov50}, see a short review in 
the Appendix A.  Linear system
(\ref{ker3}) defines the standard frame associated with the Pl\"ucker image.
The reduction equations 
(\ref{red3}), which are the compatibility conditions of (\ref{ker3}),
are nothing but the  projective `Gauss-Codazzi' equations of surfaces in $P^3$,
see  (\ref{GC1}).

All odd order flows of the DS hierarchy preserve the kernel of $D_3$ and 
the quadratic form $S$.
Hence, they induce geometric evolutions of isotropic congruences 
(and, therefore, surfaces in $P^3$), preserving the 
parametrization $x, y$ (which is conjugate in $P^5$ and asymptotic in $P^3$),
and the projectively invariant functional
$$
\int \beta \gamma \ dxdy,
$$
which  is the projective area.

%%%%%%%%%%%%%%%%%%%%%%%%%%%%%%%%%%%%%%%%%%%%%%%%%%%%%%%%%%%%%%%%%%%%%%%%%%%%%
\subsection{N-component case}
%%%%%%%%%%%%%%%%%%%%%%%%%%%%%%%%%%%%%%%%%%%%%%%%%%%%%%%%%%%%%%%%%%%%%%%%%%%%%
In the N-component case, equations (\ref{D3}) take the form
\begin{gather*}
\Psi_i^*=c_i(\partial_i^3 \Psi_i - 2 V_i\partial_i \Psi_i - 
(\partial_iV_i)\Psi_i) \\
+\sum\limits_{\substack{1\leqslant p \leqslant N,\\ p\neq i}}
c_p
(\beta_{ip}\partial_p^2\Psi_p - (\partial_p \beta_{ip})\partial_p\Psi_p
+(\partial_p^2\beta_{ip}-2V_p\beta_{ip})\Psi_p),
\end{gather*}
the corresponding reduction equations being
\begin{gather}
c_i
(\partial_i^3\beta_{ji}- 2V_i\partial_i\beta_{ji}- (\partial_i V_i)\beta_{ji})
+c_j
(\partial_j^3\beta_{ij}- 2V_j\partial_j\beta_{ij}- (\partial_j V_j)\beta_{ij}) 
\nn\\
+ \sum\limits_{\substack{1\leqslant p \leqslant N,\\ p\neq i,j}}
c_p
(\beta_{ip}\partial_p^2\beta_{jp}+\beta_{jp}\partial_p^2\beta_{ip}
-(\partial_p\beta_{ip})(\partial_p\beta_{jp})- 2V_p\beta_{ip}\beta_{jp})=0,
\label{red3N}
\\
\partial_j V_i=2\beta_{ij}\partial_i \beta_{ji}+
\beta_{ji}\partial_i \beta_{ij},\quad i\neq j.
\nn
\end{gather}
For $N=2$ they reduce to (\ref{D3}) under the identification
$\beta_{12}=\beta, \ \beta_{21}=\gamma$. The kernel of $D_3$ is defined by the linear
system
\begin{gather*}
\partial_i\Psi_j=\beta_{ij} \Psi_i,\\
\partial_i^3 \Psi_i = 2 V_i\partial_i \Psi_i + 
(\partial_iV_i)\Psi_i
\\
-\sum\limits_{\substack{1\leqslant p \leqslant N,\\ p\neq i}}
\frac{c_p}{c_i}
(\beta_{ip}\partial_p^2\Psi_p - (\partial_p \beta_{ip})\partial_p\Psi_p
+(\partial_p^2\beta_{ip}-2V_p\beta_{ip})\Psi_p),
\end{gather*}
which is compatible of rank $3N$ by virtue of the reduction equations
(\ref{red3N}). Each of the $\Psi_i$'s can thus be regarded as a position vector of an
$N$-dimensional submanifold $M_i$ in $P^{3N-1}$, parametrized by conjugate coordinates. 
Moreover, the line 
$(\Psi_i, \ \Psi_j)$ is tangential to both $M_i$ and $M_j$. 
Introducing the quadratic form $S$ by 
the equations
\bea
\partial_iS=\Psi_i\Psi_i^*,
\label{Sdiff}
\eea
one can readily show that this expression explicitly integrates to 
\bea
S=\sum_i c_i(\Psi_i\partial_i^2\Psi_i-\frac{1}{2}(\partial_i\Psi_i)^2-
V_i\Psi_i^2),
\label{SN}
\eea
thus defining the invariant symmetric scalar product on the kernel of $D_3$. One can show 
that all congruences 
$(\Psi_i, \Psi_j)$ are isotropic with respect to $S$.

To the best of our knowledge, the geometric object consisting of
N  submanifolds  in $P^{3N-1}$, $N\geq 3$,  connected  
by isotropic congruences, has not been discussed before. 
In is not clear, for instance, 
whether such structure can be related to the Pl\"ucker image of an N-dimensional 
projective submanifold carrying a 
holonomic asymptotic net. 
%%%%%%%%%%%%%%%%%%%%%%%%%%%%%%%%%%%%%%%%%%%%%%%%%%%%%%%%%%%%%%%%%%%%%%%%%%%%%%%%%
\section{Reductions in the framework of multicomponent 
KP hierarchy}
%%%%%%%%%%%%%%%%%%%%%%%%%%%%%%%%%%%%%%%%%%%%%%%%%%%%%%%%%%%%%%%%%%%%%%%%%%%%%%%
The main purposes of this section are 

--- to demonstrate that 
the reduction (\ref{reduction0})
of multicomponent KP hierarchy leads to the existence of transformations (\ref{D}),
(\ref{trans})  used to characterize a class of reductions 
in geometric context;
 
--- to  explicitly  construct the kernel of this transformations;

--- to calculate generating potentials $S$ for reductions of
arbitrary order. 

First we need to identify geometric objects
(wave functions, potentials $\beta_{ij}$) in the framework of
multicomponent KP hierarchy.

In our description of the infinite-dimensional Grassmannian we will
follow Witten \cite{Witten}, considering the spaces $W$ and  $W^*$
as dual spaces of boundary values of the operator $\dbar$,
for both of which this operator has zero index. Slightly changing the standard
setting for technical convenience, we will consider the problem
of inversion of the $\dbar$-operator in the unit disc with center at zero
(not at infinity), so that in the formula (\ref{parameterization}) and in the
expression for $F(\lambda)$ one should change $\lambda$ to $\lambda^{-1}$,
$F_{ij}=\delta_{ij}F_i(\lambda)=
\delta_{ij}c_i\lambda^{-n}$,
\bea
g_i(\lambda)=g_i(\lambda,\bbox{x})=\exp
\left(\sum_{n=1}^{\infty}x_{i(n)}\lambda^{-n}\right).
\label{param}
\eea
The case when  the operator $\dbar$
is invertible (has a kernel of zero dimension) for both spaces of boundary values, 
corresponds to the principal
Grassmannian stratum. In this case both spaces $W$ and  $W^*$ are
transversal to the space of functions analytic in the unit disc.
The evolution of a point belonging to the principal stratum
is taking place in the principal stratum almost everywhere in $\bbox{x}$,
except a manifold of codimension one,
where the objects we need for geometry (wave functions,
potentials) have singularities. So we will consider the dynamics
only on the principal stratum.

In the principal stratum, the inversion of  the operator 
$(-2\pi \text{i})^{-1}\dbar_\lambda$
with the space of boundary values $W(\lambda)$, and  the inversion of the dual
operator $2\pi \text{i}\dbar_\mu$ with the space of boundary values $W^*(\mu)$,
is defined by the same Green function (Cauchy kernel) $\chi(\lambda,\mu)$
\cite{LVBbook}, having very simple analytic properties:  it
is  an $N\times N$ matrix function analytic in both variables in $\bbox{D}$ outside
$\lambda=\mu$, behaving as $(\lambda-\mu)^{-1}$ near $\lambda=\mu$ .
An arbitrary function with these properties defines a pair of dual points
in the principal stratum 
$W$, $W^*$; the dynamics of the Cauchy kernel is characterized by Hirota's
bilinear identity, 
\bea
\oint
\chi(\nu,\mu;\bbox{x})g(\nu,\bbox{x}){g}^{-1}(\nu,\bbox{x'})
\chi(\lambda,\nu;\bbox{x'})d\nu=0, 
\label{HIROTA}
\eea 
implied by (\ref{dual}) and (\ref{dynamics})
(see \cite{LVBbook}). Thus, the Cauchy kernel gives a compact
representation of a dual pair of points on the Grassmannian, and
defines all  objects  considered in the geometric
setting.

Linear equations (\ref{Dirac}), (\ref{Diracadj}), (\ref{Darboux}),
(\ref{*Darboux}), as well as linear transformations defining the reductions
(\ref{D}),(\ref{trans}) are connected with expansions of
some functions belonging to $W$ and $W^*$ with respect to a special basis.
To construct this basis in the spaces $W$ and  $W^*$, we start with the 
functions $\chi_i(\lambda;\bbox{x})=
\chi_{i \underline~}(\lambda,0;\bbox{x})\in W(\lambda, \bbox{x})$ and
$\chi^*_i(\lambda;\bbox{x})=\chi_{\underline~i}(0,\mu;\bbox{x})\in W^*(\lambda, \bbox{x})$,
where  $\chi_{i \underline~}$ and $\chi_{\underline~ i}$
 denote $i$-th row and $i$-th column of the matrix $\chi$. Using 
the operators $\D_i$, $\D^*_i$, $1\leqslant i \leqslant N$,
\begin{gather*}
\D_i W_j(\lambda,\bbox{x})=(\partial_i +\delta_{ij}\lambda^{-1})W_j,\\
\D^*_i W^*_j(\lambda,\bbox{x})=(\partial_i -\delta_{ij}\lambda^{-1})W^*_j,
\end{gather*}
possessing the property
\begin{gather*}
\D_i W(\lambda,\bbox{x})\subset W(\lambda,\bbox{x}),\\
\D^*_i W^*(\lambda,\bbox{x})\subset W^*(\lambda,\bbox{x}),
\end{gather*}
(this property is readily checked using (\ref{dynamics}) and (\ref{param});
in the $\dbar$-dressing method the ring of operators
possessing this property is known as the Zakharov-Manakov ring \cite{ZM85}),
we obtain a basis in the form
\bea
\D_i^n\chi_i\in W,\quad \D_i^{*n}\chi^*_i\in W^*,
\quad 0\leqslant n <\infty,\quad 1\leqslant i \leqslant N.
\label{basis1}
\eea

Expressing the functions $\D_i\chi_j(\lambda,\bbox{x})\in W$ and
$\D^*_i\chi_j(\lambda,\bbox{x})\in W^*$, $i\neq j$, through this basis, we get
\bea
\D_i\chi_j(\lambda,\bbox{x})=\beta_{ij}(\bbox{x})\chi_i(\lambda,\bbox{x}),
\label{Darb1}\\
\D^*_i\chi_j(\lambda,\bbox{x})=\beta_{ji}(\bbox{x})\chi^*_i(\lambda,\bbox{x}),
\label{Darb2}
\eea
where
$$
\beta_{ij}(\bbox{x})=\chi_{ji}(0,0;\bbox{x}).
$$
In terms of Baker-Akhieser functions,
\bea
&&
\psi_{i}(\lambda,\bbox{x})= 
\chi_{i}(\lambda;\bbox{x}){g}(\lambda,\bbox{x})\in W_0(\lambda),
\nn\\&&
\psi^*_{i}(\lambda,\bbox{x})= {g}^{-1}(\lambda,\bbox{x})
\chi^*_{i}(\lambda;\bbox{x})\in W^*_0(\lambda),
\nn
\eea
the operators $\D_i$ and  $\D^*_i$ act as usual differentiations,
and (\ref{Darb1}), (\ref{Darb2}) imply
\bea
\partial_i\psi_j(\lambda,\bbox{x})=\beta_{ij}(\bbox{x})\psi_i(\lambda,\bbox{x}),
\label{Darb1a}\\
\partial_i\psi^*_j(\lambda,\bbox{x})=\beta_{ji}(\bbox{x})\psi^*_i(\lambda,\bbox{x}).
\label{Darb2a}
\eea
Introducing scalar wave functions
\bea
&&
\Psi_{i}(\bbox{x})=\oint \psi_{i}(\lambda,\bbox{x}) \rho(\lambda) d\lambda,
\label{Psi}\\
&&
\Psi^*_{i}(\bbox{x})=\oint \rho^*(\lambda)
\psi^*_{i}(\lambda,\bbox{x})  d\lambda,
\label{Psi*}
\eea
where $\rho(\lambda)$ (column) and  $\rho^*(\lambda)$ (row) 
are arbitrary weight functions, we readily derive linear equations (\ref{Darboux}) and 
(\ref{*Darboux})
from (\ref{Darb1a}) and (\ref{Darb2a}) .

Similarly, let us define a reduction by the formula 
(\ref{reduction0}) and consider the expansion of
${\chi_i^*}^{\text{t}}(-\lambda;\bbox{x})F(\lambda)\in W(\lambda,\bbox{x})$
in the basis (\ref{basis1}), 
\bea
{\chi_i^*}^{\text{t}}(-\lambda;\bbox{x})F(\lambda)=c_i \D_i^n \chi_i(\lambda,\bbox x)+
\sum_{j=1}^N\sum_{p=0}^{n-1} U_{jn}(\bbox{x})\D_j^p \chi_j(\lambda,\bbox{x}),
\nn
\eea
in terms of dual wave functions 
\bea
\Psi^{*\text{red}}_i(\bbox{x})=
\oint {\psi^*_i}^{\text{t}}(-\lambda;\bbox{x})F(\lambda)\rho(\lambda)d\lambda,
\label{wave}
\eea
which can be represented in the form (\ref{Psi*}) with
$$
\rho^*(\lambda)=\rho^{*\text{red}}(\lambda)=-\rho^{\text{t}}(-\lambda)F(-\lambda).
$$
In this way one arrives at the
linear transformations (\ref{D}) and (\ref{trans})
which were used  to define reductions in  geometric context.

Let us consider the kernel of linear transformations (\ref{D}), (\ref{trans}).
This kernel is connected with the linear space of weight functions
$\rho(\lambda)$, for which the dual wave functions 
$\Psi^{*\text{red}}_i(\bbox{x})$ are equal to zero,
\bea
\Psi^{*\text{red}}_i(\bbox{x})=
\oint {\psi^*_i}^{\text{t}}(-\lambda;\bbox{x})F(\lambda)\rho(\lambda)d\lambda=0.
\nn
\eea
It is easy to conclude that
\bea
\rho(\lambda)\in F^{-1}(\lambda) W_0^{\text{t}}(-\lambda).
\nn
\eea
Using the reduction condition (\ref{reduction0}) and some linear algebra,
we represent the linear space
$F^{-1}(\lambda) W_0^{\text{t}}(-\lambda)$ in the form
\bea
F^{-1}(\lambda) W_0^{\text{t}}(-\lambda)=
W_0^*(\lambda)\oplus {W_0^*}^{\text{an}}(\lambda),
\label{represent}
\eea
where ${W_0^*}^{\text{an}}(\lambda)$ is a finite-dimensional space of functions
analytic in the unit disc. This space has the dimension $2n$ for the transformation
(\ref{D}), and $Nn$ for the transformation (\ref{trans})
($n$ is the order of reduction, i.e., the power of $\lambda^{-1}$
in $F(\lambda)$, and $N$ is the number of components).
The basis in this space is given by the functions
\bea 
\rho^p_i(\lambda)=
F^{-1}(\lambda) \partial_i^p\psi_i^{\text{t}}(-\lambda,\bbox{x})_{\bbox{x=0}}
\in {W_0^*}^{\text{an}}(\lambda),
\label{rhokernel}
\eea
where $0\leqslant p\leqslant n-1$.
The representation (\ref{represent}) implies that
wave functions belonging to the kernel of transformations
(\ref{D}), (\ref{trans}), 
\bea
\Psi_i(\bbox{x})=\oint 
{\psi_i}(\lambda;\bbox{x})\rho(\lambda)d\lambda,
\quad \rho(\lambda)\in W_0^*(\lambda)\oplus {W_0^*}^{\text{an}}(\lambda),
\nn
\eea
correspond to the weight functions
belonging to $W_0^{\text{an}}(\lambda)$
(for $\rho(\lambda)\in W_0^*(\lambda)$ wave functions evidently equal zero),
and the dimension of the kernel coincides with the dimension of the analytic space.
An arbitrary wave function belonging to the kernel 
can be expressed as
\bea
\Psi_i(\bbox{x})=
\oint \psi_i(\lambda,\bbox{x})
\left(\sum_{k=1}^N\sum_{p=0}^{n-1}c_k^p
\rho_k^p(\lambda)\right) d\lambda,
\label{wavekernel}
\eea
where $c_k^p$ are arbitrary constants, and the functions
$\rho_i^p(\lambda)$ are given by the formula (\ref{rhokernel}).
Wave functions belonging to the 
kernel
are specified by the values of the functions
$\partial_i^p\Psi_i(\bbox{x})$, $0\leqslant p\leqslant n-1$,
$1\leqslant i\leqslant N$ at the initial point $\bbox{x=0}$
(i.e., these values define the constants in (\ref{wavekernel})). 
This is an immediate corollary of (\ref{D}), (\ref{trans}), 
or directly  (\ref{wavekernel}).
One can demonstrate that the evolution of the vector
of these values with respect to some time of the hierarchy
is defined by a closed set of linear equations, and to 
explicitly express this vector 
through the initial vector for arbitrary values of times.
%%%%%%%%%%%%%%%%%%%%%%%%%%%%%%%%%%%%%%%%%%%%%%%%%%%%%%%%%%%%%%%%%%%%%%%%%%%%%%%%%%
\subsection{Generating forms $S$ for reductions
of arbitrary order}
%%%%%%%%%%%%%%%%%%%%%%%%%%%%%%%%%%%%%%%%%%%%%%%%%%%%%%%%%%%%%%%%%%%%%%%%%%%%%%%%%%
Now we will construct generating quadratic forms $S$ 
for reductions 
of arbitrary order (see expressions (\ref{complexN})  and(\ref{SN}) 
in  second and third order cases).
Generating forms explicitly define transformations (\ref{D}),
(\ref{trans})  (in the odd order case these forms are defined
by the formula (\ref{Sdiff}), in the even order 
case this formula is replaced by (\ref{Sdiffeven})). 
An interesting feature is that for odd order
reductions the forms are symmetric, while for even order reductions 
they are antisymmetric. This  implies that reductions of 
even and odd orders are connected with geometric objects of different 
types, and an illustration of that is given by our detailed consideration 
of  reductions of first, second  and third orders.

In terms of the Cauchy kernel, the definition of reduction
(\ref{reduction0}) leads to the equation
\bea
\oint\chi^{\text{t}}(\lambda,-\nu;\bbox{x})F(\nu)\chi(\mu,\nu;\bbox{x})d\nu=0,
\quad \mu,\lambda\in \bbox{D},
\label{reduction1}
\eea
(equations of this type were used in \cite{LVBbook}
to define reductions in terms of the Cauchy kernel).
The integration of this equation gives
\bea
\chi^{\text{t}}(\lambda,-\mu;\bbox{x})F(\mu)-F(-\lambda)\chi(\mu,-\lambda;\bbox{x})=
\text{Res}_{\nu=0}\left(\chi^{\text{t}}(\lambda,-\nu;\bbox{x})F(\nu)
\chi(\mu,\nu;\bbox{x})\right).
\label{redint}
\eea
It is convenient to introduce another basis in the spaces
$W$ and $W^*$ defined through the Cauchy kernel by the expansions
\bea
&&
\chi_{i \underline~}(\lambda,\mu;\bbox{x})=
\sum_{n=0}^{\infty} \chi_{i(n)}(\lambda,\bbox{x})\mu^n, 
\quad \chi_{i(n)}(\lambda,\bbox{x})\in W(\lambda,\bbox{x}),
\label{basis2a}\\&&
\chi_{\underline~ i}(\lambda,\mu;\bbox{x})=
\sum_{n=0}^{\infty} \chi^*_{i(n)}(\mu,\bbox{x})\lambda^n,
\quad
\chi^*_{i(n)}(\mu,\bbox{x})\in W^*(\mu,\bbox{x}),
\label{basis2b}
\eea
where $0\leqslant n <\infty,\; 1\leqslant i \leqslant N$ and
$\chi_{i(0)}(\lambda,\bbox{x})=\chi_{i}(\lambda,\bbox{x})$,
$\chi^*_{i(0)}(\mu,\bbox{x})=\chi^*_{i}(\mu,\bbox{x})$ in terms of
notations introduced  before. Then it is possible to transform
the r.h.s. of equation (\ref{redint}) to
\bea
\text{Res}_{\nu=0}\left(\chi^{\text{t}}(\lambda,-\nu;\bbox{x})F(\nu)
\chi(\mu,\nu;\bbox{x})\right)=
\sum_{i=1}^{N}
c_i \sum_{p+q=n-1}
\chi^{\text{t}}_{i(p)}(\lambda,\bbox{x})(-1)^p \chi_{i(q)}(\mu,\bbox{x}).
\label{Res}
\eea
Performing a transition to scalar wave functions, 
we define the generating form $S_n$ by the expression
\bea 
&&
S_n(\bbox{x})=\sum_{i=1}^N c_i S^i_n,
\label{Sn}\\
&&
S_n^i=\frac{1}{2}\sum_{p+q=n-1}
\Psi'_{i(p)}(\bbox{x})(-1)^p \Psi_{i(q)}(\bbox{x}),
\nn
\eea
where we have introduced higher wave functions
\bea
&&
\Psi_{i(p)}(\bbox{x})= 
\oint\chi_{i(p)}(\lambda,0;\bbox{x}){g}(\lambda,\bbox{x})\rho(\lambda)
d\lambda,
\eea
$\Psi_{i(0)}(\bbox{x})=\Psi_{i}(\bbox{x})$;
for $\Psi'_{i(n)}(\bbox{x})$ the weight function is $\rho'(\lambda)$
(an arbitrary second weight function).
Higher wave fuctions can be expressed through the wave functions $\Psi_i,\Psi'_i$
and their derivatives by virtue of the formulae connecting the basises
(\ref{basis1}) and  (\ref{basis2a}),
\bea
&&
\Psi_{i(p+1)}=\partial_i \Psi_{i(p)}-\beta_{ii}^p\Psi_{i},
\label{connection1}\\
&&
\partial_j \Psi_{i(p)}=\beta_{ji}^p\Psi_j,\quad i\neq j,\quad 0\leqslant p<\infty.
\label{connection2}
\eea
Here $\beta_{ij}^0=\beta_{ij}$, and higher potentials $\beta_{ii}^p,\beta_{ji}^p$
are connected with $\beta_{ij}$ by the relations implied by compatibility
conditions for the relations (\ref{connection1}) and
(\ref{connection2}),
\bea
&&
\beta_{ji}^{p+1}=\partial_i\beta_{ji}^{p} - \beta_{ji}\beta_{ii}^p,
\label{conn1}\\
&&
\partial_j\beta_{ii}^p=\beta_{ij}\beta_{ji}^p.
\label{conn2}
\eea
Using  relations (\ref{connection1}) recursively, it is possible to express
$S_n$ as a quadratic form in $\Psi_i,\Psi'_i$ and their derivatives, with
the coefficients connected with $\beta_{ij}$ by the formulae (\ref{conn1}) and
(\ref{conn2}) (and thus to get expressions of the type (\ref{complexN}) and (\ref{SN})). 
It is easy
to see that the form (\ref{Sn}) is symmetric for odd $n$ and antisymmetric for 
even $n$, i.e., for odd $n$ it is invariant under the permutation
of $\Psi_i$ and $\Psi'_i$, and for even $n$ the permutation changes the sign
of the expression (the permutation of wave functions corresponds to the permutation
of weight functions $\rho,\rho'$).

To obtain the formula of the type (\ref{Sdiffeven}) and (\ref{Sdiff}),
we will use the  general equation
\bea
\partial_i \Phi(\bbox{x})=\Psi_i(\bbox{x})\Psi^*_i(\bbox{x}),
\label{Phi}
\eea
where $\Psi_i(\bbox{x})$, $\Psi^*_i(\bbox{x})$ are given
by the expressions (\ref{Psi}), (\ref{Psi*}), and
$$
\Phi(\bbox{x})=\oint \rho^*(\mu)g^{-1}(\mu,\bbox{x})\chi(\lambda,\mu;\bbox{x})
g(\lambda,\bbox{x})\rho(\lambda) d\lambda d\mu,
$$
which follows directly from Hirota's bilinear identity (\ref{HIROTA})
(see, e.g., \cite{LVBbook}).

For odd $n$, from the formulae (\ref{redint}), (\ref{Res}), (\ref{Sn})
and the equation (\ref{Phi}) we obtain
\bea
\partial_i S_n(\bbox{x})=
\frac{1}{2}(\Psi_i{\Psi^{*\text{red}}_i}'+\Psi'_i\Psi^{*\text{red}}_i),
\label{Sdiff1}
\eea
where $\Psi^{*\text{red}}_i$ is defined by (\ref{wave}) .
Taking $\rho=\rho'$, we get exactly the formula (\ref{Sdiff}).
Thus, in the odd order case, the
reduction is characterized by the equation (\ref{Sdiff1}), where the
generating form is given by the expression (\ref{Sn}). 
Using (\ref{conn1}) and (\ref{conn2}) one can verify
that partial derivatives $\partial_j$
of each of the terms $S^i_n$ in the expression (\ref{Sn}) 
can be represented  as
$$
\partial_i S^j_n=\frac{1}{2}(\Psi_i (D^j_n \Psi')_i+\Psi'_i (D^j_n \Psi)_i),
$$
where $D^j_n$ are linear differential operators.
Then from the equation (\ref{Sdiff1}) we obtain the explicit formula
for the transformations (\ref{D}) and (\ref{trans}),
\bea
\Psi^{*\text{red}}_i=\sum_{j=1}^{N}c_j(D^j_n \Psi)_i.
\nn
\eea

In the even order case instead of (\ref{Sdiff1}) we have
\bea
\partial_i S_n(\bbox{x})=
\frac{1}{2}(\Psi_i{\Psi^{*\text{red}}_i}'-\Psi'_i\Psi^{*\text{red}}_i)=
\Psi_i\wedge\Psi^{*\text{red}}_i.
\label{Sdiff2}
\eea
Partial derivatives $\partial_j$
of the terms $S^i_n$ in the expression (\ref{Sn}) can be expressed as
$$
\partial_i S^j_n=\Psi_i\wedge (D^j_n\Psi)_i.
$$
The function $\Psi^{*\text{red}}_i$ 
(respectively, transformations (\ref{D}), (\ref{trans})),
are defined by the equation (\ref{Sdiff2})  up to a
term $U_i(\bbox{x})\Psi_i$,
\bea
\Psi^{*\text{red}}_i=U_i(\bbox{x})\Psi_i+\sum_{j=1}^{N}c_j(D^j_n \Psi)_i,
\label{trans2n}
\eea
which can be fixed either by taking Baker-Akhiezer
functions as wave functions and considering analytic properties in $\lambda$,
or just by a direct substitution into the dual operator. Nevertheless,
the {\em existence} of transformations (\ref{D}), (\ref{trans})
(given by the formula (\ref{trans2n})),
is implied by the equation (\ref{Sdiff2}), and  can be used
as a definition of the reduction.

%%%%%%%%%%%%%%%%%%%%%%%%%%%%%%%%%%%%%%%%%%%%%%%%%%%%%%%%%%%%%%%%%%%%%%%%%%%%%%%%%
\section{Appendix A: surfaces in projective differential geometry}
%%%%%%%%%%%%%%%%%%%%%%%%%%%%%%%%%%%%%%%%%%%%%%%%%%%%%%%%%%%%%%%%%%%%%%%%%%%%%%%%

Projective differential geometry of surfaces $M^2$ in $P^3$ has been
extensively
developed in the first half of the 19th century in the works of Wilczynski, Fubini,
\^Cech, Cartan,  Tzitzeica, Demoulin, Rozet, Godeaux, Lane, Eisenhart, Finikov,
Bol and many others.
Based on \cite{Wilczynski} (see also \cite{Finikov37}, \cite{Fer1}, \cite{Fer3}), 
let us briefly recall
Wilczynski's approach to the construction of  surfaces $M^2$
in projective space $P^3$ in terms of solutions of a linear system
\begin{equation}
\begin{array}{c}
{\bf r}_{xx}=\beta \ {\bf r}_y+\frac{1}{2}(V-\beta_y) \ {\bf r} \\
\ \\
{\bf r}_{yy}=\gamma \ {\bf r}_x+\frac{1}{2}(W-\gamma_x) \ {\bf r}
\end{array}
\label{r}
\end{equation}
where $\beta, \gamma, V, W$ are functions of $x$ and  $y$. If we
cross-differentiate (\ref{r}) and assume
${\bf r}, {\bf r}_x, {\bf r}_y, {\bf r}_{xy}$ to be independent, we arrive at
the compatibility
conditions \cite[p.\ 120]{Lane}
\begin{equation}
\begin{array}{c}
\beta_{yyy}-2\beta_yW-\beta W_y=
\gamma_{xxx}-2\gamma_xV-\gamma V_x\\
\ \\
W_x=2\gamma \beta_y+\beta \gamma_y \\
\ \\
V_y=2\beta \gamma_x+\gamma \beta_x,
\end{array}
\label{GC1}
\end{equation}
which  coincide with (\ref{red3}). For any fixed
$\beta, \gamma, V, W$ satisfying (\ref{GC1}), the linear system
(\ref{r})  is compatible
and possesses a solution
${\bf r}=(r^0, r^1, r^2, r^3)$ where $r^i(x, y)$
 can be regarded as homogeneous coordinates of a
surface in projective space $P^3$. One may think of
$M^2$ as a surface in a three-dimensional  space with position vector
${\bf R}=(r^1/r^0, r^2/r^0, r^3/r^0)$.
If we choose any other solution
$\tilde{{\bf r}}=(\tilde r^0, \tilde r^1, \tilde r^2, \tilde r^3)$
of the same system
(\ref{r}) then the corresponding surface
$\tilde M^2$ with position vector
$\tilde{{\bf R}}=(\tilde r^1/\tilde r^0, \tilde r^2/\tilde r^0,
\tilde r^3/\tilde r^0)$
constitutes a projective transform of $M^2$, so that any fixed
$\beta, \gamma, V, W$ satisfying (\ref{GC1}) define a surface $M^2$ uniquely up
to
projective equivalence. Moreover, a simple calculation yields
\begin{equation}
\begin{array}{c}
{\bf R}_{xx}=\beta \ {\bf R}_y+ a \ {\bf R}_x \\
{\bf R}_{yy}=\gamma \ {\bf R}_x+ b \ {\bf R}_y
\end{array}
\label{affine}
\end{equation}
$(a=-2 r^0_x/r^0,\,b=-2 r^0_y/r^0)$ which implies that $x, y$ are asymptotic
coordinates on $M^2$. System (\ref{affine}) can be viewed
as an "affine gauge" of system (\ref{r}).
In what follows, we assume that our surfaces are
hyperbolic and the corresponding asymptotic coordinates $x, y$ are
real.\footnote{The elliptic case is dealt with in an analogous manner by
regarding $x,y$ as complex conjugates.}
Since equations~(\ref{GC1})
specify a surface uniquely up to projective equivalence, they can be viewed as
the `Gauss-Codazzi' equations in projective geometry.

Even though the coefficients $\beta, \gamma, V, W$ define a surface $M^2$
uniquely up to projective equivalence via (\ref{r}),
it is not entirely correct
to regard $\beta, \gamma, V, W$ as projective invariants. Indeed, the
asymptotic
coordinates $x, y$ are only defined up to an arbitrary reparametrization
of the form
\begin{equation}
x^*=f(x), ~~~~ y^*=g(y)
\label{newxy}
\end{equation}
which induces a scaling of the surface vector according to
\begin{equation}
{\bf r}^*=\sqrt {f'(x)g'(y)}~{\bf r}.
\label{newr}
\end{equation}
Thus \cite[p.\ 1]{Bol}, the form of
equations (\ref{r}) is preserved by the above transformation
with the new coefficients $\beta^*, \gamma^*, V^*, W^*$ given by
\begin{equation}
\begin{array}{c}
\beta^{*}=\beta g'/(f')^2, ~~~~  V^{*}(f')^2=V+S(f)\\
\ \\
\gamma^{*}=\gamma f'/(g')^2, ~~~~ W^{*}(g')^2=W+S(g),
\end{array}
\label{new}
\end{equation}
where $S(\,\cdot\,)$ is the  Schwarzian derivative, that is,
$$
S(f)=\frac{f'''}{f'} - \frac{3}{2} \left(\frac{f''}{f'}\right)^2.
$$
The transformation formulae (\ref{new}) imply that the symmetric 2-form
$$
2 \beta \gamma\,dxdy
$$
and the conformal class of the cubic form
$$
\beta \,dx^3+\gamma \,dy^3
$$
are absolute projective invariants. They are known as the projective metric and
the Darboux cubic form, respectively, and play an important role in projective
differential geometry. In particular, they define a `generic' surface
uniquely up to projective equivalence. The vanishing of the Darboux cubic form
is characteristic for quadrics: indeed, in this case $\beta = \gamma =0$ so that
asymptotic curves of both families are straight lines. The vanishing of the
projective metric (which is equivalent to either $\beta =0$ or $\gamma =0$)
characterises ruled surfaces. In what follows we exclude these two degenerate
situations and require $\beta \ne 0$, $\gamma \ne 0$.

Using (\ref{newxy})-(\ref{new}), one can  verify that the four
points
\begin{equation}
\begin{array}{c}
{\bf r}, ~~~
{\bf r}_1={\bf r}_x-\frac{1}{2}\frac{\gamma_x}{\gamma}{\bf r}, ~~~
{\bf r}_2={\bf r}_y-\frac{1}{2}\frac{\beta_y}{\beta}{\bf r}, \\
\ \\
{\mbox{\boldmath $\eta$}}={\bf r}_{xy}-\frac{1}{2}\frac{\gamma_x}{\gamma}{\bf
r}_y-
\frac{1}{2}\frac{\beta_y}{\beta}{\bf r}_x+
\left(\frac{1}{4}\frac{\beta_y\gamma_x}{\beta \gamma} -
\frac{1}{2}{\beta \gamma}\right){\bf r} \\
\end{array}
\label{frame}
\end{equation}
are defined in an invariant way, that is, under the transformation formulae
\mbox{(\ref{newxy})-(\ref{new})}
they acquire a nonzero multiple which does not
change them as points in projective space $P^3$. These points form the
vertices of the so-called Wilczynski moving tetrahedron \cite{Bol},
\cite{Finikov37}, \cite{Wilczynski}.
Since the lines passing through ${\bf r}, {\bf r}_1$ and ${\bf r}, {\bf r}_2$
are
tangential to the $x$- and $y$-asymptotic curves,
respectively, the three points ${\bf r}, {\bf r}_1, {\bf r}_2$
span the tangent plane of the surface $M^2$.
The line through ${\bf r}_1, {\bf r}_2$ lying in the tangent
plane is
known as the directrix of Wilczynski of the second kind. The line through
${\bf r}, {\mbox{\boldmath $\eta$}}$ is transversal to $M^2$ and is known as
the directrix of
Wilczynski of the first kind. It plays the role of a projective `normal'.
Wilczynski's tetrahedron proves to
be the most convenient tool in projective differential geometry.

Using (\ref{r}) and~(\ref{frame}),
we easily derive for ${\bf r}, {\bf r}_1, {\bf r}_2, {\mbox{\boldmath $\eta$}}$
the linear equations \cite[p.\ 42]{Finikov37}
\begin{equation}
\begin{array}{c}
\left(\begin{array}{c}
{\bf r}\\
{\bf r}_1\\
{\bf r}_2\\
{\mbox{\boldmath $\eta$}}
\end{array}\right)_x=
\left(\begin{array}{cccc}
\frac{1}{2}\frac{\gamma_x}{\gamma}& 1&0&0\\
\frac{1}{2}b & -\frac{1}{2}\frac{\gamma_x}{\gamma}& \beta&0\\
\frac{1}{2}k&0&\frac{1}{2}\frac{\gamma_x}{\gamma}&1\\
\frac{1}{2}\beta
a&\frac{1}{2}k&\frac{1}{2}b&-\frac{1}{2}\frac{\gamma_x}{\gamma}
\end{array}\right)
\left(\begin{array}{c}
{\bf r}\\
{\bf r}_1\\
{\bf r}_2\\
{\mbox{\boldmath $\eta$}}
\end{array}\right)\\
\ \\
\left(\begin{array}{c}
{\bf r}\\
{\bf r}_1\\
{\bf r}_2\\
{\mbox{\boldmath $\eta$}}
\end{array}\right)_y=
\left(\begin{array}{cccc}
\frac{1}{2}\frac{\beta_y}{\beta}& 0&1&0\\
\frac{1}{2}l & \frac{1}{2}\frac{\beta_y}{\beta}& 0&1\\
\frac{1}{2}a&\gamma &-\frac{1}{2}\frac{\beta_y}{\beta}&0\\
\frac{1}{2}\gamma b&\frac{1}{2}a
&\frac{1}{2}l&-\frac{1}{2}\frac{\beta_y}{\beta}
\end{array}\right)\left(
\begin{array}{c}
{\bf r}\\
{\bf r}_1\\
{\bf r}_2\\
{\mbox{\boldmath $\eta$}}
\end{array}\right),
\end{array}
\label{Wilczynski}
\end{equation}
where we introduced the notation
\begin{equation}
\begin{array}{c}
k=\beta \gamma - (\ln \beta)_{xy}, ~~~~ l=\beta \gamma - (\ln \gamma)_{xy},\\
\ \\
a=W-(\ln \beta)_{yy}-\frac{1}{2}(\ln \beta)_y^2, ~~~~
b=V-(\ln \gamma)_{xx}-\frac{1}{2}(\ln \gamma)_x^2.
\end{array}
\label{klab}
\end{equation}
The compatibility conditions of equations (\ref{Wilczynski}) imply
\begin{equation}
\begin{array}{c}
(\ln \beta)_{xy}=\beta \gamma -k, ~~~~ (\ln \gamma)_{xy}=\beta \gamma -l,\\
\ \\
a_x=k_y+\frac{\beta_y}{\beta}k, ~~~~ b_y=l_x+\frac{\gamma_x}{\gamma}l,\\
\ \\
\beta a_y+2a\beta_y=\gamma b_x+2b\gamma_x,
\end{array}
\label{GC2}
\end{equation}
which is just the equivalent form of the projective "Gauss-Codazzi" equations
(\ref{GC1}).

Equations (\ref{Wilczynski}) can be rewritten in the Pl\"ucker coordinates.
For a convenience of the reader we briefly recall this construction.
Let us consider a line $l$ in $P^3$ passing through the points ${\bf a}$ and
${\bf b}$  with the
homogeneous coordinates ${\bf a}=(a^0:a^1:a^2:a^3)$ and ${\bf
b}=(b^0:b^1:b^2:b^3)$. With
the line $l$ we associate a point ${\bf a}\wedge {\bf b}$ in projective space
$P^5$ with
the homogeneous coordinates
$$
{\bf a}\wedge {\bf b}=(p_{01}:p_{02}:p_{03}:p_{23}:p_{31}:p_{12}),
$$
where
$$
p_{ij}=\det
\left(\begin{array}{cc}
a^i & a^j \\
b^i & b^j
\end{array}\right).
$$
The coordinates $p_{ij}$ satisfy the well-known quadratic Pl\"ucker relation
\begin{equation}
p_{01}\, p_{23}+p_{02}\, p_{31}+p_{03}\, p_{12}=0.
\label{quadric}
\end{equation}
Instead of ${\bf a}$ and ${\bf b}$ we may consider an arbitrary linear
combinations thereof
without changing ${\bf a}\wedge{\bf b}$ as a point in $P^5$.
Hence, we arrive at the well-defined
Pl\"ucker coorrespondence $l({\bf a},{\bf b})\to {\bf a}\wedge {\bf b}$
between lines in
$P^3$ and points on the Pl\"ucker quadric in $P^5$.
Pl\"ucker correspondence plays an important role in the projective
differential geometry of surfaces and often sheds some new light on those
properties of surfaces which are not `visible' in $P^3$ but acquire a precise
geometric meaning only in $P^5$. Thus, let us consider a surface
$M^2\in P^3$ with the Wilczynski tetrahedron
${\bf r}, {\bf r}_1, {\bf r}_2, {\mbox{\boldmath $\eta$}} $
satisfying equations (\ref{Wilczynski}).
Since the two pairs of points ${\bf r}, {\bf r}_1$ and ${\bf r}, {\bf r}_2$
generate two lines in $P^3$ which are tangential
to the $x$- and $y$-asymptotic curves, respectively, the formulae
$$
{\cal U}={\bf r} \wedge {\bf r}_1,\quad {\cal V}={\bf r} \wedge {\bf r}_2
$$
define the images of these lines under the Pl\"ucker embedding. Hence, with any
surface $M^2\in P^3$ there are canonically associated two surfaces
${\cal U}(x, y)$ and ${\cal V}(x, y)$ in $P^5$ lying on the
Pl\"ucker quadric (\ref{quadric}).
In view of the formulae
$$
{\cal U}_x=\beta \, {\cal V},\quad {\cal V}_y=\gamma \, {\cal U},
$$
we conclude that the line in $P^5$ passing through a pair of points
$({\cal U}, {\cal V})$ can also be generated by the pair of points
$({\cal U}, {\cal U}_x)$ (and hence is
tangential to the $x$-coordinate line on the surface ${\cal U}$) or by a pair
of
points $({\cal V}, {\cal V}_y)$ (and hence is tangential to the $y$-coordinate
line on the
surface ${\cal V}$).
Consequently, the surfaces ${\cal U}$ and ${\cal V}$ are two focal surfaces of
the congruence
of straight
lines $({\cal U}, {\cal V})$ or, equivalently,
${\cal V}$ is the Laplace transform of ${\cal U}$ with
respect to $x$ and ${\cal U}$ is the
Laplace transform of ${\cal V}$ with respect to $y$.

Introducing
$$
\begin{array}{c}
{\cal A} ={\bf r}_2\wedge {\bf r}_1+{\bf r}\wedge {\mbox{\boldmath $\eta$}},
~~~
{\cal B} ={\bf r}_1\wedge {\bf r}_2+{\bf r}\wedge {\mbox{\boldmath $\eta$}}, \\
\ \\
{\cal P} = 2\, {\bf r}_2\wedge {\mbox{\boldmath $\eta$}}, ~~~
{\cal Q} = 2\, {\bf r}_1\wedge {\mbox{\boldmath $\eta$}},
\end{array}
$$
we arrive at the following equations for the Pl\"ucker coordinates:

\begin{equation}
\begin{array}{c}
\left(\begin{array}{c}
{\cal U}\\
{\cal A}\\
{\cal P}\\
{\cal V}\\
{\cal B}\\
{\cal Q}
\end{array}\right)_x=
\left(\begin{array}{cccccc}
0 & 0 & 0 & \beta & 0 & 0\\
k & 0 & 0 & 0 & 0 & 0\\
0 & k & 0 & -\beta a & 0 & 0\\
0 & 0 & 0 & \frac{\gamma_x}{\gamma} & 1 & 0\\
0 & 0 & 0 & b & 0 & 1\\
-\beta a & 0 & \beta & 0 & b &-\frac{\gamma_x}{\gamma}
\end{array}\right)
\left(\begin{array}{c}
{\cal U}\\
{\cal A}\\
{\cal P}\\
{\cal V}\\
{\cal B}\\
{\cal Q}
\end{array}\right)\\
\ \\
\left(\begin{array}{c}
{\cal U}\\
{\cal A}\\
{\cal P}\\
{\cal V}\\
{\cal B}\\
{\cal Q}
\end{array}\right)_y=
\left(\begin{array}{cccccc}
\frac{\beta_y}{\beta} & 1 & 0 & 0 & 0 & 0\\
a & 0 & 1 & 0 & 0 & 0\\
0 & a & -\frac{\beta_y}{\beta} & -\gamma b & 0 & \gamma\\
\gamma & 0 & 0 & 0 & 0 & 0\\
0 & 0 & 0 & l & 0 & 0\\
-\gamma b & 0 & 0 & 0 & l & 0
\end{array}\right)
\left(\begin{array}{c}
{\cal U}\\
{\cal A}\\
{\cal P}\\
{\cal V}\\
{\cal B}\\
{\cal Q}
\end{array}\right)
\end{array}
\label{UAPVBQ}
\end{equation}
Equations (\ref{UAPVBQ}) are consistent with the following table
of scalar products:
\begin{equation}
({\cal U}, {\cal P})=-1, ~~~ ({\cal A}, {\cal A})=1, ~~~
({\cal V}, {\cal Q})=1, ~~~ ({\cal B}, {\cal B})=-1,
\label{table}
\end{equation}
all other scalar products being equal to zero. This defines a scalar
product of the signature (3, 3) which is the same as that of the quadratic form
(\ref{quadric}). Equivalently, one can say that the quadratic form
$$
S={\cal Q} \ {\cal V}-{\cal P} \ {\cal U}+\frac{{\cal A}^2-{\cal B}^2}{2},
$$
is an integral of (\ref{UAPVBQ}). The explicit form of $S$ in terms of 
${\cal U}$ and ${\cal V}$
is
$$
S={\cal V}_{xx}{\cal V}-\frac{1}{2}{\cal V}_x^2-V{\cal V}^2-
({\cal U}_{yy}{\cal U}
-\frac{1}{2}{\cal U}_y^2
-W{\cal U}^2).
$$
Notice that equations (\ref{UAPVBQ}) and the expression for $S$
identically coincide with (\ref{ker3}) and (\ref{S}) if one sets
${\cal U}=\Psi_2, \ {\cal V}=\Psi_1, \ c_1=-c_2=1$.

\section{Appendix B: congruences W}

There exists an important class of transformations in projective differential
geometry which leave the system (\ref{r}) form-invariant. These 
transformations are
generated by congruences W, and require a solution of  certain 
Dirac equation on the surface $M^2$. Here we briefly recall this construction following
\cite{Jonas1}, \cite{Finikov37}, \cite{Eisenhart}.

Let $M^2$ be a surface with the position vector ${\bf r}$ satisfying (\ref{r}).
With any pair of functions ${\cal U}$ and ${\cal V}$ solving the Dirac equation
(\ref{Dirac}), 
\begin{equation}
\begin{array}{c}
{\cal U}_x=\beta \ {\cal V}, \\
{\cal V}_y=\gamma \ {\cal U},
\end{array}
\label{W}
\end{equation}
we associate a surface $\tilde M^2$ with the position vector
${\bf r}'$ given by the formula
\begin{equation}
{\bf r}'={\cal V} \ {\bf r}_1- {\cal U} \ {\bf r}_2+
\frac{1}{2}\left( {\cal V}
 \ \frac{\gamma_x}{\gamma} -
{\cal U} \ \frac{\beta_y}{\beta}-{\cal V}_x+{\cal U}_y\right) ~ {\bf r}
\label{newtilder}
\end{equation}
In order to write down the equations for ${\bf r}'$, it is convenient to
introduce certain quantities which are combinations of
${\cal U}, {\cal V}$ and their derivatives. First of all, we define
${\cal A}$ and  ${\cal B}$ by the formulae
$$
{\cal U}_y=\frac{\beta_y}{\beta} \ {\cal U}+{\cal A}, ~~~~
{\cal V}_x=\frac{\gamma_x}{\gamma} \ {\cal V}+{\cal B},
$$
(in fact, we are copying equations (\ref{UAPVBQ}) for the Pl\"ucker
coordinates).
The compatibility conditions ${\cal U}_{xy}={\cal U}_{yx}$ and
${\cal V}_{xy}={\cal V}_{yx}$ imply
$$
{\cal A}_x = k \ {\cal U}, ~~~~ {\cal B}_y=l \ {\cal V},
$$
where $l$ and $k$ are the same as in (\ref{klab}). Let us introduce
${\cal P}$ and ${\cal Q}$ by the formulae
$$
{\cal A}_y=a \ {\cal U}+{\cal P}, ~~~~ {\cal B}_x=b \ {\cal V}+{\cal Q}.
$$
Then  compatibility conditions imply
$$
{\cal P}_x=-\beta a \ {\cal V}+k \ {\cal A}, ~~~~
{\cal Q}_y=-\gamma b \ {\cal U}+l \ {\cal B}.
$$
Finally, we introduce the quantities $H$ and $K$ via
$$
{\cal P}_y=a \ {\cal A}-\frac{\beta_y}{\beta} \ {\cal P} -\gamma b \ {\cal V}+
\gamma \ {\cal Q}-K, ~~~~
{\cal Q}_x=b \ {\cal B}-\frac{\gamma_x}{\gamma} \ {\cal Q} -\beta a \ {\cal U}+
\beta \ {\cal P}+H,
$$
so that compatibility conditions imply that $H$ and $K$ satisfy the equation 
dual to (\ref{W}),
\begin{equation}
H_y=\beta \ K, ~~~~ K_x=\gamma \ H.
\label{Diracdual}
\end{equation}
Equations for ${\cal U}, {\cal A}, {\cal P}, {\cal V}, {\cal B}, {\cal Q}, H,
K$
can be rewritten in matrix form
\begin{equation}
\begin{array}{c}
\left(\begin{array}{c}
{\cal U}\\
{\cal A}\\
{\cal P}\\
{\cal V}\\
{\cal B}\\
{\cal Q}\\
H\\
K
\end{array}\right)_x=
\left(\begin{array}{cccccccc}
0 & 0 & 0 & \beta & 0 & 0 & 0 & 0\\
k & 0 & 0 & 0 & 0 & 0 & 0 & 0\\
0 & k & 0 & -\beta a & 0 & 0 & 0 & 0\\
0 & 0 & 0 & \frac{\gamma_x}{\gamma} & 1 & 0 & 0 & 0\\
0 & 0 & 0 & b & 0 & 1 & 0 & 0\\
-\beta a & 0 & \beta & 0 & b &-\frac{\gamma_x}{\gamma} & 1 & 0\\
{ * } & * & * & * & * & * & * & * \\
0 & 0 & 0 & 0 & 0 & 0 & \gamma & 0
\end{array}\right)
\left(\begin{array}{c}
{\cal U}\\
{\cal A}\\
{\cal P}\\
{\cal V}\\
{\cal B}\\
{\cal Q}\\
H\\
K
\end{array}\right)\\
\ \\
\left(\begin{array}{c}
{\cal U}\\
{\cal A}\\
{\cal P}\\
{\cal V}\\
{\cal B}\\
{\cal Q}\\
H\\
K
\end{array}\right)_y=
\left(\begin{array}{cccccccc}
\frac{\beta_y}{\beta} & 1 & 0 & 0 & 0 & 0 & 0 & 0\\
a & 0 & 1 & 0 & 0 & 0 & 0 & 0\\
0 & a & -\frac{\beta_y}{\beta} & -\gamma b & 0 & \gamma & 0 & -1\\
\gamma & 0 & 0 & 0 & 0 & 0 & 0 & 0\\
0 & 0 & 0 & l & 0 & 0 & 0 & 0\\
-\gamma b & 0 & 0 & 0 & l & 0 & 0 & 0\\
 0 & 0 & 0 & 0 & 0 & 0 & 0 & \beta\\
{ * } & * & * & * & * & * & * & *
\end{array}\right)
\left(\begin{array}{c}
{\cal U}\\
{\cal A}\\
{\cal P}\\
{\cal V}\\
{\cal B}\\
{\cal Q}\\
H\\
K
\end{array}\right)
\end{array}
\label{W1}
\end{equation}
where the elements $*$ are not yet specified. Equations (\ref{W1}) reduce to
(\ref{UAPVBQ}) under the reduction $H=K=0$.
In what follows we will also need the quantity
\begin{equation}
S={\cal Q} \ {\cal V}-{\cal P} \ {\cal U}+\frac{{\cal A}^2-{\cal B}^2}{2},
\label{S1}
\end{equation}
which, in view of (\ref{W1}), satisfies the equations
\begin{equation}
S_x={\cal V} \ H, ~~~~ S_y={\cal U} \ K.
\label{S2}
\end{equation}

{\bf Remark.} The explicit form of $S$ in terms of ${\cal U}$ and ${\cal V}$
is
$$
S={\cal V}_{xx}{\cal V}-\frac{1}{2}{\cal V}_x^2-V{\cal V}^2-
({\cal U}_{yy}{\cal U}
-\frac{1}{2}{\cal U}_y^2
-W{\cal U}^2),
$$
so that
$$
\begin{array}{c}
H={\cal V}_{xxx}-2V{\cal V}_x-V_x{\cal V}-\beta {\cal U}_{yy}+\beta_y{\cal U}_y 
+(2\beta W-\beta_{yy}){\cal U}, \\
\ \\
-K={\cal U}_{yyy}-2W{\cal U}_y-W_y{\cal U}-\gamma {\cal V}_{xx}+\gamma_x{\cal V}_x 
+(2\gamma V-\gamma_{xx}){\cal V}.
\end{array}
$$
Notice that equations $H=K=0$ identically coincide with (\ref{D3}) under the
obvious identifications. In particular, 
$H$ and $K$ solve the adjoint linear problem (\ref{Diracdual}).

\bigskip

Now a direct calculation gives:
\begin{equation}
{\bf r}=-2 \frac{{\cal V}}{S} \  {\bf r}'_x-
2 \frac{{\cal U}}{S}\ {\bf r}'_y+\frac{1}{S}
({\cal A}+{\cal B}+\frac{\gamma_x}{\gamma}{\cal V}+\frac{\beta_y}{\beta}{\cal
U})
\ {\bf r}'.
\label{oldr}
\end{equation}
Equations (\ref{newtilder}) and (\ref{oldr}) imply that the line
${\bf r}\wedge  {\bf r}'$ joining the corresponding points
${\bf r}$ and ${\bf r}'$ is tangential to both surfaces $M^2$ amd $\tilde M^2$,
which are thus focal surfaces of the line congruence
${\bf r}\wedge  {\bf r}'$. Moreover, the formulae
\begin{equation}
\begin{array}{c}
{\bf r}'_{xx}=\frac{S_x}{S}\  {\bf r}'_x+
\left (\frac{S_x}{S} \frac{{\cal U}}{{\cal V}}-\beta \right)\ {\bf r}'_y+
\frac{1}{2}\left( V+\beta_y-\frac{S_x}{S{\cal V}}
({\cal A}+{\cal B}+\frac{\gamma_x}{\gamma}{\cal V}+\frac{\beta_y}{\beta}{\cal
U})
\right) {\bf r}', \\
\ \\
{\bf r}'_{yy}=\frac{S_y}{S}\  {\bf r}'_y+
\left (\frac{S_y}{S} \frac{{\cal V}}{{\cal U}}-\gamma \right)\  {\bf r}'_x+
\frac{1}{2}\left( W+\gamma_x-\frac{S_y}{S{\cal U}}
({\cal A}+{\cal B}+\frac{\gamma_x}{\gamma}{\cal V}+\frac{\beta_y}{\beta}{\cal
U})
\right)  {\bf r}',
\end{array}
\label{W2}
\end{equation}
(which are the result of quite a long calculation)
demonstrate that $x$ and $y$ are asymptotic coordinates on the transformed surface
$\tilde M^2$ as well, so that the congruence ${\bf r}\wedge {\bf r}'$
preserves the asymptotic parametrization of it's focal surfaces. Such
congruences play
a central role in projective differential geometry and are known as
 congruences W. By a construction, a congruence W with one given focal
surface
$M^2$ is uniquely determined by a solution ${\cal U}, {\cal V}$ of the linear
Dirac equation (\ref{W}). Normalising the vector ${\bf r}'$ as 
${\bf r}'={\sqrt S}~\tilde {\bf r}$, we can rewrite equations (\ref{W2}) in the
canonical form (\ref{r}),
\begin{equation}
\begin{array}{c}
\tilde {\bf r}_{xx}=\tilde \beta \ \tilde {\bf r}_y+
\frac{1}{2}(\tilde V-\tilde \beta_y) \ \tilde {\bf r} \\
\ \\
\tilde {\bf r}_{yy}=\tilde \gamma \ \tilde {\bf r}_x+
\frac{1}{2}(\tilde W-\tilde \gamma_x) \ \tilde {\bf r}
\end{array}
\label{tilder}
\end{equation}
where the transformed coefficients $\tilde \beta, \ \tilde \gamma, \ \tilde V,
\
\tilde W$ are given by the formulae
\begin{equation}
\begin{array}{c}
\tilde \beta=\frac{S_x}{S} \frac{{\cal U}}{{\cal V}}-\beta=
 \frac{H {\cal U}}{S}-\beta, ~~~~~~
\tilde \gamma=\frac{S_y}{S} \frac{{\cal V}}{{\cal U}}-\gamma=
 \frac{K {\cal V}}{S}-\gamma,
 \\
\ \\
\tilde V = V-\frac{S_x}{S} \frac{{\cal V}_x}{{\cal V}}
+\frac{3}{2}(\frac{S_x}{S})^2-\frac{S_{xx}}{S}, ~~~~~
\tilde W = W-\frac{S_y}{S} \frac{{\cal U}_y}{{\cal U}}
+\frac{3}{2}(\frac{S_y}{S})^2-\frac{S_{yy}}{S};
\end{array}
\label{W3}
\end{equation}
we point out the simple identity
$\tilde \beta \tilde \gamma = \beta \gamma - (\ln S)_{xy}$.
The surface $\tilde M^2$ is called a W-transform of the surface $M^2$.
The construction of  W-congruences on the transformed surface  
$\tilde M^2$ 
requires a solution of the transformed Dirac equation
\begin{equation}
\partial_x{\tilde \Psi}_2=\tilde \beta {\tilde \Psi}_1, ~~~~ 
\partial_y{\tilde \Psi}_1=\tilde \gamma {\tilde \Psi}_2,
\label{Diracnew}
\end{equation}
where $\tilde \beta$ and $\tilde \gamma$ are given by (\ref{W3}). Let us introduce
a potential $M$ by the formulae
$$
M_x=H\Psi_1, ~~~~ M_y=K\Psi_2,
$$
the compatibility of which readily follows from (\ref{Dirac}) and 
(\ref{Diracdual}). Then an arbitrary solution $\Psi_1, \ \Psi_2$
of (\ref{Dirac}) generates
a solution $\tilde \Psi_1, \ \tilde \Psi_2$ of (\ref{Diracnew}) by the formula
\begin{equation}
\tilde \Psi_1=\Psi_1-\frac{{\cal V} M}{S}, ~~~~
\tilde \Psi_2=\Psi_2-\frac{{\cal U} M}{S},
\label{Darbu}
\end{equation}
which is a  specialization of the Darboux-Levi transformation, 
see \cite{Schief}.

Congruences W provide a standard tool for constructing
B\"acklund transformations. Suppose we are given a
class of surfaces specified by  certain  extra constraints imposed on
$\beta, \ \gamma, \ V, \ W$.
Let us try to find a congruence W such that
the second focal surface will also belong to the same class.
This requirement imposes
additional restrictions on  ${\cal U}$ and ${\cal V}$,
which usually turn to be linear  and, moreover, contain an arbitrary
constant parameter, so that equations (\ref{W1})
become a "Lax pair" for the class of surfaces under study.
Since the Dirac equation (\ref{W}) is a part of this Lax pair, it is not
surprising
that surfaces in projective differential geometry are closely related to the DS hierarchy.
Particularly interesting classes of surfaces correspond to
reductions  which are quite familiar in the modern
soliton theory. These are

isothermally-asymptotic surfaces ($\beta=\gamma$);

surfaces $R_0$ ($\beta=1$ or $\gamma=1$),

surfaces R ($\beta_y=\gamma_x$);

surfaces of Jonas ($\beta_x=\gamma_y$), etc.

The construction of the corresponding B\"acklund transformations
was  carried out primarily by Jonas \cite{Jonas1}, see also
\cite{Fer1}, \cite{Fer3}.

%%%%%%%%%%%%%%%%%%%%%%%%%%%%%%%%%%%%%%%%%%%%%%%%%%%%%%%%%%%%%%%%%%%%
\subsection*{Acknowledgements}
We thank the London Mathematical Society for their financial support
of LVB visit to Loughborough, making this collaboration possible.
LVB was also partially supported by RFBR grant 01-01-00929.
%%%%%%%%%%%%%%%%%%%%%%%%%%%%%%%%%%%%%%%%%%%%%%%%%%%%%%%%%%%%%%%%%%%%


\begin{thebibliography}{99}
\addcontentsline{toc}{section}{References}

\bibitem{Blaschke} Blaschke~W. 
\textit{Vorlesungen \"uber Differentialgeometrie, V.3.}
Springer-Verlag, Berlin, 1929.

\bibitem{Bogdanov} Bogdanov~L.V. 
Veselov-Novikov equation as a natural
two-dimensional generalization of the Korteweg-de Vries equation, 
\textit{Theor. and Math. Phys.} 
\textbf{70} (1987) 309--314.

\bibitem{LVBbook} Bogdanov~L.V. 
\textit{Analytic-Bilinear Approach 
to Integrable Hierarchies.} 
Kluwer, Dordrecht, 1999.

\bibitem{Bol} Bol~G. 
\textit{Projektive Differentialgeometrie.}
G\"ottingen (1954).

\bibitem{Burstall} Burstall F., Pedit F. and Pinkall U. 
Schwarzian derivatives 
and flows on surfaces, \texttt{arXiv:math.DG/0111169}.

\bibitem{Eisenhart} Eisenhart L.P. 
\textit{Transformations of surfaces.} 
Chelsea Publishing Company, New York, 1962.

\bibitem{Fer1}  Ferapontov~E.V. and Schief~W.K. Surfaces of Demoulin:
  Differential geometry, B\"acklund transformation and Integrability,
  {\sl Preprint SFB 288 No.\ 316}, Berlin (1998).


\bibitem{Fer2} Ferapontov~E.V. Stationary Veselov-Novikov equation
  and isothermally asymptotic surfaces in projective differential geometry,
  {\sl Preprint SFB 288 No.\ 318}, Berlin (1998) (\texttt{DG/9805011}).

\bibitem{Fer3}  Ferapontov~E.V. 
Integrable systems in projective differential 
geometry, 
\textit{Kyushu J. Math.} \textbf{54} (2000) 183-215.

\bibitem{Fer4}  Ferapontov~E.V. 
Lie sphere geometry and integrable systems,
\textit{Tohoku Math. J.} 
\textbf{52} 
(2000) 199-233.

\bibitem{Fer5} Ferapontov~E.V. 
Surfaces with flat normal bundle: an explicit construction,
\textit{Diff. Geom. Appl.} 
\textbf{14}, N1 (2001) 15-37.

\bibitem{Finikov37} Finikov~S.P.
\textit{Projective Differential Geometry.}
Moscow-Leningrad, 1937.

\bibitem{Finikov50} Finikov~S.P. 
\textit{Theory of congruences.} 
Moscow-Leningrad,
1950.

\bibitem{Fubini} Fubini~G. and \^Cech~E. 
\textit{Geometria Proiettiva Differenziale.}
Bologna: Zanichelli, 1926.

\bibitem{GM} Grinevich~P.G. and Manakov~S.V.
Inverse problem of scattering theory for the
two-dimensional Schr\"odinger operator, the $\bar\partial$-method and nonlinear
equations. (Russian) 
\textit{Funktsional. Anal. i Prilozhen.} 
\textbf{20} (1986), no. 2, 14--24. 

\bibitem{BKP} Jimbo~M. and Miwa~T. 
Solitons and infinite-dimensional Lie algebras.
\textit{Publ. Res. Inst. Math. Sci.} 
\textbf{19} (1983), no. 3, 943--1001.

\bibitem{Jonas1} Jonas~H. \"Uber die konstruktion der W-kongruenzen zu
einem gegebenen Brennfl\"achenmantel und \"uber die Transformationen der
R-fl\"achen, Jahresber. 
\textit{Deutsch. Math. Ver.} 
\textbf{29} (1920) 40--74.

\bibitem{Kac}
Kac~V.G. and van de Leur~J.W.
The $n$-component KP hierarchy and representation theory,
in: 
Fokas, A. S. (ed.) et al., 
\textit{Important developments in soliton theory.} 
Berlin: Springer-Verlag, 302-343 (1993).

\bibitem{Kon1} Konopelchenko~B.G. Induced surfaces and their
integrable dynamics,
\textit{Studies in Appl. Math.}, 
\textbf{96}(1) 9--51 (1996).

\bibitem{Kon2} Konopelchenko~B.G. Nets in $R^3$, their
integrable evolutions and
the DS hierarchy, 
\textit{Phys. letters A} 
\textbf{183} (2-3), 153--159 (1993).

\bibitem{KonPin} Konopelchenko~B.G. and Pinkall~U. Integrable deformations of
affine surfaces via Nizhnik-Veselov-Novikov equation,
 {\sl Preprint SFB 288 No.\ 318}, Berlin (1998).


%\bibitem{Kupershmidt} Kupershmidt B. A., Comm. Math. Phys. 99 (1985) 51.

\bibitem{Lane} Lane~E. 
\textit{Treatise on Projective Differential Geometry.}
The Univ.\ of Chicago Press (1942).

%\bibitem{Lelieuvre} Lelieuvre~M., Sur les lignes asymptotiques et leur
%repr\'esentation sph\'erique,
%Bull.  Sci. Mathematiques 12 (1888) 126--128.

\bibitem{Matveev} Matveev~V.B. and Salle~M.A.
\textit{Darboux transformations and Solitons.} 
Springer Verlag, 1991.

\bibitem{Mikhailov} Mikhailov A.V. The reduction problem and the inverse 
scattering method,
\textit{Physica D} 
\textbf{3} (1981) 73--117.


%\bibitem{Nizhnik} Nizhnik~L.P., Integration of multidimensional nonlinear
%equations by the method of inverse problem, DAN SSSR, 254 (1980) 332.

\bibitem{Sato} Sato~M., Soliton equations as dynamical systems on an infinite 
dimensional 
Grassmann manifolds.
\textit{RIMS Kokyuroku} 
\textbf{439}, 30--46 (1981),\\
Sato~M. and Sato~Y.,
Soliton equations as dynamical systems on infinite dimensional Grassmann manifold. 
\textit{North-Holland Math. Stud.} \textbf{81}, 259--271 
(1983).

\bibitem{Schief} Schief W.K. On a $(2+1)$-dimensional Darboux system: 
integrable reductions, 
\textit{Inverse Problems} 
\textbf{10}  (1994), no. 5, 1185--1198. 

\bibitem{Segal}
Segal~G. and Wilson~G.
Loop groups and equations of KdV type. 
\textit{Publ. Math., Inst. Hautes \'Etud. Sci.} \textbf{61}, 5-65 (1985),\\
Pressley~A. and Segal~G.
\textit{Loop groups.} 
Oxford: Clarendon Press (1986). 

\bibitem{Taimanov} Taimanov~I.A. Modified Novikov-Veselov equation and
differential geometry of surfaces, in 
\textit{Solitons, Geometry and Topology}
(eds. V.M. Buchstaber and S.P. Novikov) 
\textit{Transl. AMS, ser.2} 
\textbf{179} (1997),
133--155.

\bibitem{VesNov} Veselov A.P. and Novikov S.P. Finite-gap two-dimensional
potential Schr\"odinger operators. Explicit formulae and evolution
equations, 
\textit{DAN SSSR}, 
\textbf{279}(1) (1984) 20--24.

\bibitem{Wilczynski} Wilczynski E.I. Projective-differential geometry of
curved surfaces, 
\textit{Trans. AMS} 
\textbf{8} (1907) 233--260; \textbf{9} (1908) 79--120, 293--315.

\bibitem{Wilczynski1} Wilczynski E.I.  Sur la th\'eorie g\'en\'erale des congruences,
M\'emoire couronn\'e par la classe des sciences. 
M\'emoires publi\'es par la Classe des Sciences de l'Acad\'emie Royale de Belgique. 
Collection en 4. Deuxi\'eme s\'erie. Tome III (1911). 

\bibitem{Wilczynski2} Wilczynski E.I. The general theory of congruences, 
\textit{Trans. Amer. Math. Soc.}  
\textbf{18}  (1915)  311--327.

\bibitem{ZM85}
Zakharov~V.E. and Manakov~S.V.
Construction of higher-dimensional nonlinear integrable systems and of their solutions. 
\textit{Funct. Anal. Appl.} \textbf{19}, 89--101 (1985).


\bibitem{Zakharov1} Zakharov V. E. 
Description of the $n$-orthogonal curvilinear coordinate systems and 
Hamiltonian integrable systems of hydrodynamic type. I. 
Integration of the Lam\'e equations,
\textit{Duke Math. J.} 
\textbf{94}  (1998), no. 1, 103--139.

\bibitem{ZM} Zakharov~V.E. and Manakov~S.V.
On reduction in systems integrable by method of inverse problem dispersion,
\textit{DAN}, \textbf{360}(3) (1998), 324--327.

\bibitem{Zakharov2} Zakharov V. E.
Integration of the Gauss-Codazzi equations,
\textit{Teoret. Mat. Fiz.} \textbf{128}  (2001), no. 1, 133--144. 

\bibitem{Witten}Witten~E. Quantum field theory, Grassmannians,
and algebraic curves,
\textit{Comm. Math. Phys.} 
\textbf{18}(4) (1988), 529--600.

\end{thebibliography}
\end{document}